\documentstyle[12pt, epsf]{article}
\begin{document}
\begin{center}
{ \Large \bf 
%%%%%%%%%%%%
On Bose-Einstein condensate
%%%%%%%%%%%%%%
}
\end{center}
\begin{center}
{ \Large \bf  
%%%%%%%%%%%%
inside moving exciton-phonon droplets
%%%%%%%%%%%%%%
}
\end{center}
%%%%%%%%%%%%%%%%%%%%%%%%%%%%%%%%
\begin{center} 
\vspace{2mm}
 \vspace*{5mm}
{\large \bf D. Roubtsov\,$^1$,  Y.  L\'epine}
\end{center} 
%%%%%%%%%%%%%%%%%%%%%%%%
\begin{center}
 \medskip
{ \it 
GCM et D\'epartement de Physique, Universit\'e de Montr\'eal 
}  
 \end{center}
\begin{center}
{\it 
C.P. 6128,  Succ. Centre-ville,  Montr\'eal, Qu\'ebec, Canada H3C 3J7
}
\end{center}
\vspace*{5mm}
%%%%%%%%%%%%%%%%%%%%%%%%%%%%%%%%%
\begin{center} 
{\large \bf I. Loutsenko\,$^2$}
\end{center} 
%%%%%%%%%%%%%%%%%%%%%%%%%%%
\begin{center} 
{\it 
Jadwin Hall, Physics Department, Princeton University
}
\end{center} 
\begin{center} 
{\it 
Princeton,  New Jersey, USA 08544
} 
\end{center}
%%%%%%
%% 
%% \maketitle
\vspace*{5mm}
\begin{quotation}
We explore a nonlinear field model to describe the interplay 
between  
the ability of  excitons to be  Bose condensed  
and   
their interaction  with other modes of a crystal. 
We apply our consideration to the long-living 
paraexcitons in Cu$_{2}$O.
Taking into account the exciton-phonon interaction and introducing   
a coherent phonon part of the moving condensate, we solve 
quasi-stationary
equations for the exciton-phonon condensate. 
These equations support localized  solutions,
and  we discuss the conditions
for the inhomogeneous condensate to appear in the crystal.
Allowable values of  the characteristic width of 
ballistic condensates are estimated in the limit $T \rightarrow 0$.   
The stability conditions of the moving condensate are analyzed by use of
Landau arguments, and Landau critical parameters appear in the theory.
It follows that, under certain conditions, 
the condensate can move through the crystal
as a stable droplet. 
To separate the coherent and  
non-coherent parts of the exciton-phonon packet, 
we suggest to turn off
the phonon wind by the changes in design of the 3D crystal and 
boundary conditions  for the  moving droplet.   

\end{quotation}

\vspace*{0.5cm}


\newline
\hspace*{2.5 cm} Exciton-Phonon condensate; Phonon Wind; Stability
 
\vspace*{0.5cm}
 
$^1$\,E-mail: roubtsod@physcn.umontreal.ca

$^2$\,E-mail: loutseni@feynman.princeton.edu

\newpage

\section{Introduction}

Nowadays, there is a lot of experimental evidence that  
paraexcitons in
Cu$_{2}$O crystals can form a strongly correlated state, 
which can be assigned to the excitonic  Bose Einstein condensate 
(BEC) \cite{Andre},\cite{Lin},\cite{Goto}.

Surprisingly enough, a cloud of excitons that  seems to contain the condensate
can be prepared in a moving state, and such a packet 
moves ballistically through the crystal at $T < T_{c}$. 
At $T>T_{c}$, however, 
the excitonic packet exhibits the standard diffusive behavior.
Thus,  in the there dimensional Cu$_{2}$O crystals,
the excitonic Bose Einstein condensate is found to be in 
a spatially inhomogeneous  state
with the well defined characteristic width $L_{\rm ch}$ in the direction of
motion \cite{Andre}. 
The registered ballistic velocities of the coherent exciton packets 
turn out to be always less, but  approximately equal to 
the longitudinal sound speed of the crystal, \,$v \,<\,c_{\rm s}$.
Note that paraexcitons in the pure Cu$_{2}$0 crystals have the extremely large
lifetime,  $\tau\simeq 13\,\mu$s, 
and, at the conditions we discuss in this work,  they
are optically inactive \cite{book}. 
Although transferring of  the moving  coherent excitonic field into 
a coherent photon field  is not a hopeless task \cite{Fernandez},\cite{PS},
no convincing experimental results are obtained yet, \cite{Lin},\cite{Butov}.
%%%%%%%%%%%%%%
%%
%%%%%%%%%%

To understand  the physics of 
anomalous excitonic transport,  we accept \cite{book},\cite{LR},\cite{RL} 
that  the macroscopic wave 
function $\Psi_{0}\sim \phi_{\rm o}\,e^{i\varphi_{\rm c}}$ can be
associated with the coherent part of the excitonic packet.
(Here $\varphi_{\rm c}$ is the coherent phase of the condensate.)
In other words, the experimental results  \cite{Andre}
suggest the following decomposition
of the density of excitons in the packet,  
\begin{equation}
n({\bf x},t)=n_{\rm coh}({\bf x},t) + \Delta n({\bf x},t), 
\end{equation} 
where  $n_{\rm coh}({\bf x},t) \approx n_{\rm coh}(x -vt) $
is the ballistic ({\it superfluid}) part of the packet, 
\begin{equation}
n_{\rm coh}(x -vt) \simeq  \vert \Psi_{0}\vert^{2}(x-vt),  
\end{equation} 
and $ \Delta n({\bf x},t)$ is the diffusive (non-condensed) part of it, 
\begin{equation}
 \Delta n({\bf x},t) \simeq \langle  \delta\hat{\psi}^{\dag} 
\delta\hat{\psi}  ({\bf x},t) \rangle. 
\end{equation} 
Therefore, the problem is how to describe a spatially inhomogeneous     
state of the excitonic BEC in terms of $\Psi_{0}({\bf x},t)$ and   
$\delta\hat{\psi}  ({\bf x},t)$, where $\delta\hat{\psi}$ is the ``fluctuating'' 
part of the exciton Bose field.  
Indeed, if a coherent excitonic packet moves  in a crystal (or another
semiconductor 
structure), it interacts with phonons, non-condensed excitons, 
impurities and other imperfections of the lattice, etc., 
(see Ref. \cite{Lozovik}).

All this makes 
the problem of superfluidity of the Bose-condensed excitons  
a rather complicated and challenging one. 

In this Letter,   
we start from the simplest possible approximation:
we describe the moving condensate only and show that    
a sort of Gross Pitaevskii equation 
\cite{Griffin},\cite{Pit}
do has sense  at $T\ll T_{c}$.
In this case,  the finite characteristic length of the condensate
(but not of the total exciton-phonon packet)
appears naturally
in the framework of the effective 1D nonlinear Schr\"odinger equation, 
by which we model the real 3D conditions.

\section{Exciton-Phonon Condensate}

To obtain the necessary density of excitons in the excitonic cloud and, thus,  meet the BEC
conditions, the  Cu$_{2}$0 crystals were irradiated  
by laser pulses with 
$\hbar \omega_{L} \gg E_{\rm gap}$ at $T\simeq 2\sim 5$\,K.
Note  
that the cross-section area of an excitation spot on a surface of the crystal, $S$,
can be made large enough, so that  $S \simeq S_{\rm surf}$. 
Then, the classical ``phonon wind'', 
or the flow of nonequilibrium phonons from the surface into the bulk \cite{ST2}, 
can transfer the nonzero momentum to the excitonic cloud,  
${\bf P}_{\rm exc} \simeq N_{\rm x}\,\langle \hbar\, {\bf k}_{0}\rangle \ne 0$ \,and\,  
${\bf P}_{\rm exc} \perp S_{\rm surf}$.

As a result, the packet of moving excitons and nonequilibrium phonons 
of the phonon wind ($N_{\rm ph}\simeq N_{\rm x}$)
is actually the system that undergoes the transition toward the Bose Einstein condensation.
Then, one can estimate the energy of the packet without any condensate as follows 
$$
E \simeq N_{\rm x}\bigl\{(\hbar^{2}/2m_{\rm x})\langle
{\bf k}_{0}^{2}\rangle +  
2\nu_{0} n_{\rm x} \bigr\}+
N_{\rm ph}\,\langle\hbar \omega_{\rm ac}'\rangle, 
$$
where $\hbar \langle k_{0\,x} \rangle = m_{\rm x}
\langle v \rangle  \simeq m_{\rm x}\,c_{\rm s}$ and 
$\nu_{0}>0$
is the exciton-exciton interaction strength, and 
$n_{\rm x}$ is the average density of excitons in the packet.

At $T <T_{c}(n_{\rm x})$,  or, equivalently,  $n_{\rm x} > n_{c}(T)$ \cite{book}, 
the condensate can be formed inside the excitonic  droplet and 
the following representation of the exciton Bose-field holds,  
$\hat{\psi}=\Psi_{0} + \delta\hat{\psi}$.
(As the kinetics of condensate formation is not a subject 
of this paper, we assume  $T_{\rm cloud} \simeq T$ and $T\rightarrow 0$.)
Moreover,   
for the  displacement field of the crystal, 
%% $\hat{\bf u}$,  
we can introduce 
a nontrivial coherent part too, i.e. $\hat{\bf u}={\bf u}_{0} + \delta\hat{\bf u}$ and
${\bf u}_{0} \ne 0$. Then,  
a moving coherent packet can contain the exciton-phonon condensate,  
which is the self-consistent
exciton-phonon field, 
$\Psi_{0}({\bf x},t) \cdot {\bf u}_{0}({\bf x},t)$.  

The macroscopic wave function of excitons,  \,$\Psi_{0}({\bf x},t) \approx \Psi_{0}(x,t)$,  
is normalized as follows 
\begin{equation} 
\int\!\!\vert\Psi_{0}\vert^{2}(x,t)\,d{\bf x}
=S\! \int\!\!\phi_{\rm o}^{2}(x)\,dx = N_{\rm o},
\label{norma} 
\end{equation}
where $N_{\rm o}$ is the (macroscopic) number of condensed excitons, and,
generally, 
\,$N_{\rm o} \ne N_{\rm x}$. 
To model the ballistic motion of $n_{\rm coh}(x,t)$, we use  
the following {\it ansatz},  
\begin{equation}
\Psi_{0}(x,t) =
{\rm e}^{-i (\tilde{E}_{g}\,+ \,m_{\rm x}v^{2}/2 \,-\,\vert{\mu}\vert  )\,t }
\,{\rm e}^{ i(\varphi + k_{0}x)}\,
\phi_{\rm o}(x-vt),
\label{ansatz1}
\end{equation}
%%%%%%%%%%%%%%%%%%%%%%%
\begin{equation}
 u_{0\,j}(x,t)=u_{\rm o}(x-vt)\,\delta_{1j},
\label{ansatz2}
\end{equation}
where \,$\tilde{E}_{g}= E_{\rm gap}-E_{\rm x}$, 
$E_{\rm x}$ is the exciton Rydberg, 
$\varphi={\rm const}$,
\,$\hbar k_{0}=m_{\rm x} v$, 
and $\mu$ is 
the effective chemical potential of the condensate.  
Note that there is no difference between the ballistic velocity of the 
exciton-phonon packet
and the superfluid velocity of the condensate 
if we start from ansatz (\ref{ansatz1}).

For the envelope function $\phi_{\rm o}(x)$ and the phonon direct current
$u_{\rm o}(x)$, 
we obtain 
the following  stationary equations \cite{LR},\cite{RL} 
%%%%%
%%%%
\begin{equation} 
-\vert{\mu}\vert\,\phi_{\rm o}(x)=\bigl( -(\hbar^{2}/2m_{\rm x}) 
\partial_{x}^{2} \,+\, \tilde{\nu}_{0}\,\phi_{\rm o}^{2}(x)
\, + \,
\tilde{\nu}_{1}\,\phi_{\rm o}^{4}(x)\,\bigr)\,\phi_{\rm o}(x), 
\label{1Deq}
 \end{equation}
%%%%%%%%%%%%%%%%%%%%
\begin{equation} 
\partial_{x}u_{\rm o}(x) \approx {\rm const}_{1}\,\phi_{\rm o}^{2}(x) +
{\rm const}_{2}\, \phi_{\rm o}^{4}(x).
\label{1Deq1} 
\end{equation}

Here, the ``bare''  exciton  vertices, 
both the two-particle $\nu_{0}>0$ and the three-particle $\nu_{1}>0$, 
can be strongly  
renormalized because of exciton-phonon
interaction.   
We choose it in the simplest form of  Deformation Potential,
\begin{equation}
\tilde{E}_{g}\,\hat{\psi}^{\dag}\hat{\psi} \,\rightarrow\,
(\tilde{E}_{g}\, + \,\sigma_{0}\,\partial_{j} \hat{u}_{j})\,\hat{\psi}^{\dag}\hat{\psi},\,
\,\,\,\,\, \sigma_{0}>0. 
\label{DP}
\end{equation}
Yet the first (cubic) anharmonicity $\kappa_{3} \ne 0$ 
is taken into account to model the lattice.
Note that we consider the case $v \simeq c_{\rm s}$, and, at first glance,  
the (adiabatic) assumption 
of coherent propagation of the crystal deformation field 
and the excitonic condensate  might not be valid. 
However, the nonlinear lattices support some localized excitations 
that can consist of 
two parts, namely, the  direct current and alternating one, 
$$
\hat{u} \simeq u_{\rm o}(x-\tilde{v}t) + \delta \hat{u}(x-\tilde{v}t,\,t).
$$
Such excitations 
can move with the (group) velocity $\tilde{v} \simeq c_{\rm s}$ \cite{Hu}.
%%%%%%%%%%%%%
%%%%%%%%%%%%%%%%%%%%%%%%%\cite 
Therefore,  Eq. (\ref{1Deq1}) can be used to describe the exciton-phonon
condensate with $v \simeq c_{\rm s}$.

If the following conditions 
\begin{equation}
\tilde{\nu}_{0} = \tilde{\nu}_{0}(\,\nu_{0},\, \sigma_{0}, \,v/c_{\rm s}\,)< 0
\,\,\,\,\,{\rm and}\,\,\,\,\, 
\tilde{\nu}_{1}= \tilde{\nu}_{1}(\,\nu_{1},\, \sigma_{0}, \,v/c_{\rm s}, \,\kappa_{3} \,) >0
\label{Cond}
\end{equation} 
can be valid,    
the localized (solitonic) solution of Eq. (\ref{1Deq}) exists. It 
can be written in the following form:
%%%%%%
% Note: the following equations in two lines if necessary
%%%%%%%%
%% $$
%% \phi_{\rm o}(x)=\Phi_{\rm o}\,f(\beta\bigl(\Phi_{\rm o}\bigr)\,x), 
%% $$
%% $$
%% \beta(\Phi_{\rm o})=\sqrt{(2m_{\rm x}/\hbar^{2})\,\vert\mu\vert\bigl(\Phi_{\rm
%% o}\bigr)\,}, 
%% $$
\begin{equation}
\phi_{\rm o}(x)=\Phi_{\rm o}\,f(\beta\bigl(\Phi_{\rm o}\bigr)x,
\,{\rm const}\bigl(\Phi_{\rm o}\bigr)\,), 
\,\,\,\,\,\,\,\,
\beta(\Phi_{\rm o})=\sqrt{(2m_{\rm x}/\hbar^{2})\,\vert\mu\vert\bigl(\Phi_{\rm
 o}\bigr)\,}, 
\label{LOC}
\end{equation}
where $\Phi_{\rm o}$ is the amplitude of the soliton and  
$1/\beta(\Phi_{\rm o})\equiv L_{0}$ is its width. Then,     
the characteristic 
length of the condensate can be estimated as 
$$
L_{\rm ch} \simeq (2\sim 4)\,L_{0} \propto \vert \mu \vert ^{-1/2}, 
$$
and 
the effective chemical potential of the condensate, $\vert\mu\vert (N_{\rm o},\,v)$,
defines the value of $L_{\rm ch}$ in this model.

We  can use the simplest approximation,  
\begin{equation}
\vert\mu\vert(\Phi_{\rm o}) \approx  \vert \tilde{\nu}_{0}\vert \,\Phi_{\rm o}^{2}/2
\,\,\,\,\,{\rm and}\,\,\,\,\,\vert\mu\vert(N_{\rm o},\,v) \propto N_{\rm o}^{2}, 
\label{mu}
\end{equation}
which turns out to be valid at 
$
\bar{n}_{\rm o} > 5 - 10
$.
Here \,$\bar{n}_{\rm o}=N_{\rm o}^{*}/N_{\rm o}$, \,and\,  
$ N_{\rm o}^{*}= 2\,S /{a}_{\rm x}^{2}$ \,is 
a macroscopically large parameter ($a_{\rm x}$ is the exciton Bohr radius),
$$
N_{\rm o}^{*} \simeq 10^{13} \sim 10^{14}.
$$
Indeed, 
one can estimate  $\vert \mu \vert$ more  accurately,
\begin{equation}
\vert\mu\vert(\Phi_{\rm o})  = \vert \tilde{\nu}_{0}\vert \,\Phi_{\rm o}^{2}/2\,-\, 
 \tilde{\nu}_{1}\,\Phi_{\rm o}^{4}/3
\,\,\,\,\,{\rm and}\,\,\,\,\,
\vert\mu\vert(N_{\rm o},\,v) \propto  \Bigl(2 \,\bar{n}_{\rm o}^{2}\,{\rm x}\,E_{\rm x}\,a_{\rm x}^{6}
\,+\,
4\,\tilde{\nu}_{1}(v/c_{\rm s})\,\Bigr)^{-1}. 
\end{equation}
Here, ${\rm x}=\mu_{\rm x}/m_{\rm x}$ and the last estimate is valid for $N_{\rm o}< N_{\rm o}^{*}$.

We choose  approximation (\ref{mu}), 
and the  characteristic length of the condensate
can be 
estimated as follows (\,$\tilde{\nu}_{0} = 
\tilde{\varepsilon}_{0}(v/c_{\rm s}) \,{a}_{\rm x}^{3}$ \,\,and\,\, $\nu_{0} = 
\varepsilon_{0}\,{a}_{\rm x}^{3}$\,):
\begin{equation}
L_{\rm ch}(N_{\rm o},\,v) \simeq 
4\,
\sqrt{ \frac{ \hbar^{2}}{ m_{\rm x}\,\vert \tilde{\nu}_{0}\vert }\,\Phi_{\rm o}^{-2}\bigl(N_{\rm o},\,v\bigr)\,}
\,\simeq \,
4 \,\frac{E_{\rm x}}{\vert\tilde{\varepsilon}_{0}(v/c_{\rm s})\vert}\,\bar{n}_{\rm o}\,
a_{\rm x}\, \propto\, N_{\rm o}^{-1}. 
\label{estimate1}
\end{equation}

At $v=v_{\rm o}$, or, equivalently, at $\gamma(v)=\gamma_{\rm o}$,  where
$$
\gamma(v)=c_{\rm s}^{2}/\bigl(c_{\rm s}^{2}-v^{2}\bigr) \,\,\,\,{\rm and}\,\,\,\, v<c_{\rm s},
$$
we \,have $\tilde{\varepsilon}_{0}(v_{\rm o}/c_{\rm s})=0$ \,and\, $\gamma_{\rm o} \simeq 3\sim 5$ \cite{LR}.
Therefore,  solitonic solution (\ref{LOC}) and estimate (\ref{estimate1}) are valid at 
\,$v_{\rm o}< v < c_{\rm s}$, or \,$\gamma(v)>\gamma_{\rm o}$.
For example, 
at
$\gamma(v) \approx 2\,\gamma_{\rm o}\simeq 6 \sim 10$,
(\,$\gamma(v=0.95\,c_{\rm s}) \approx 10$\,), we obtain \,$\tilde{\nu}_{0}<0$\, and 
estimate \,$\tilde{\varepsilon}_{0}(v/c_{\rm s}) \approx - \varepsilon_{0}$.

Thus, at \,$\gamma(v)>\gamma_{\rm o}$ \,and\,  $\bar{n}_{\rm o} > 10$ 
(for estimates we take $\vert\tilde{\varepsilon}_{0}(v/c_{\rm s})\vert \simeq (10^{-2}\sim 10^{-1})\,E_{\rm x}$ 
  and  $\bar{n}_{\rm o} \simeq 10^{1}\sim 10^{2}$, i.e. 
$N_{\rm o} \simeq 10^{11} \sim 10^{12}$),
we obtain  
a large factor multiplied by $a_{\rm x}$ as an estimate of $L_{\rm ch}$ \cite{RL}, 
e.g., 
$$
L_{\rm ch} = {\cal F}\,a_{\rm x} \gg a_{\rm cr}.
$$
Here  
$a_{\rm cr}\simeq 4$\,{\AA}  is the lattice constant and 
$a_{\rm x}^{3} \simeq (3\sim 4)\,a_{\rm cr}^{3}$ for the paraexcitons in Cu$_{2}$O.

Within the quasi-stationary approximation 
(\ref{ansatz1})-(\ref{1Deq}) at $T=0$, 
this is a physically reasonable result, ${\cal F}\simeq 10^{2} \sim 10^{4}$,  
and the duration of the  condensate can be estimated as   
$t_{\rm ch} \simeq 2 \cdot (10^{-11} - 10^{-9})$\,s.
%%%%%%%%
Although these results are in a qualitative agreement 
with the average 3D densities of 
the Bose-condensed excitons in moving packets at $T\ne 0$,
$n_{\rm o}\simeq 10^{17}\sim 10^{18}$\,cm$^{-3}$ \cite{Andre}, 
%%%%%%%%%%%%%%%%%%%%%%%%
%%%%%%%%%%%%%%%%%%%%%%%%%%%%%%%
the characteristic duration of such localized packets is 
$\Delta t \simeq (4 \sim 8)\cdot 10^{-7}$\,s experimentally.

\section{Stability against Outside Excitations}

Analyzing experimental results \cite{Andre}, one can notice that 
a long-lasting ``tail'' of excitons is followed by  
the coherent (localized) excitonic packet.  
This tail might  be explained
by instability of the exciton-phonon condensate moving in 
the lattice. Indeed, such an instability can lead to continuous emission of 
the excitons  out from the condensate. As $\partial_{t}N_{\rm o}< 0$,  
we have  $\partial_{t}L_{\rm ch}(t) > 0$ and, as a result, 
nonstationary transport of the ballistic condensate.  
Alternatively, the condensate is stable, and diffusive propagation of the 
non-condensed excitons, $\Delta N \simeq  N_{\rm x} - N_{\rm o} $, 
is responsible for the tail in $n(x,t)$.
%%%%
%% 
We argue that the second  
scenario seems to be true.
%%%%%%%

For outside collective excitations (see Fig. 1), the asymptotics of the low-lying  
energy spectrum can be found easily.  Indeed, if we assume that       
$\phi_{\rm o}^{2}(x) \approx 0$ 
and $\partial_{x}u_{\rm o}(x) \approx 0$
in the outside packet area, the excitonic and phonon branches
are (formally) uncoupled. 
Then, for the excitonic branch,  we obtain in the co-moving frame
\begin{equation}
\hbar\omega_{{\rm ex}}({\bf k})\approx \vert\mu\vert + (\hbar^{2}/2m)k^{2},\,\,\,\,\, 
{\rm u}_{\bf k}({\bf x})\approx u_{\bf k}\,{\rm e}^{i{\bf k}{\bf x}},\,\,\,
{\rm v}_{\bf k}({\bf x})\approx 0,\,\,\,\,\, \vert x\vert \gg L_{0}, 
\label{gap1}
\end{equation}
where ${\rm u}_{\bf k}$ and  ${\rm v}_{\bf k}$ are Bogoliubov-deGennes amplitudes,
and \,$\omega_{\rm ph}(\tilde{\bf k})=c_{\rm s}\vert \tilde{\bf k} \vert $ in 
the laboratory frame of reference. 
Note that  the  condition \,$\hbar\omega_{\rm ex}({\bf k}) + \hbar\,k_{x}\,v  > 0$ can be violated
at the velocities close to $v_{\rm o}$. 
This is the hint that the instability regime can occur \cite{LR}.

We assume that the exciton and phonon can be emitted from the condensate coherently.  
Then the emission of $\delta N$ excitons out of the coherent packet can 
be described  as an appearance of $\delta N$ outside collective excitations 
in the system condensate plus medium, see Fig. 1. 
Therefore,  one can use Landau arguments to analyze the stability of the condensate. 
It turns out that  the moving condensate is stable against the direct emission of outside excitations.
However, this is not the case for the inside excitations. 
In the laboratory frame, these excitations can be represented in the form 
$$
\delta\psi ({\bf x}, t) \sim 
{\rm u}_{k}(x-vt)\,{\rm e}^{ i(\varphi_{0} +{\bf k}{\bf x})} \,{\rm e}^{-i(\omega_{k}+k_{x}v)t } +
{\rm v}_{k}(x-vt)\,{\rm e}^{-i(\varphi_{0} + {\bf k}{\bf x})}\,{\rm e}^{i(\omega_{k}+k_{x}v)t},  
$$
and 
\begin{equation}
\delta u ({\bf x}, t) \sim 
C_{k}(x-vt)\,\exp\bigl(i(\varphi_{0}+ {\bf k}{\bf x})\bigr)\,\exp\bigl(-i(\omega_{k}+k_{x}v)\,t \bigr)+ {\rm c.c.}, 
\label{urfin1}
\end{equation}
and the following asymptotic behavior is valid for the Bogoliubov-deGennes amplitudes 
(\,$\phi_{\rm o}(x/L_{0}) \sim \exp(-\vert x \vert/L_{0} )$\,):
$$
{\rm u}_{k}(x) \sim {\rm v}_{k}(x) \sim \phi_{\rm o}(x/L_{0}), 
\,\,\,\,\,
C_{k}(x) \sim \phi_{\rm o}^{2}(x/L_{0})
\,\,\,\,{\rm at}\,\,\,\, \vert x \vert > L_{0}.
$$
We can start from  the standard criterion, 
\begin{equation}
\hbar (\omega^{(-)}_{ {\rm  ex},\,k} -   
\vert  k_{x} \vert  v) >0 \,\,\,\,\,{\rm at }\,\,\,\,\,\,\vert  k_{x} \vert \simeq  z\,L_{0}^{-1},
\label{STable}
\end{equation}   
where  the superscript $(-)$ means $k_{x}<0$ and 
$z \simeq  3 \sim 10$ corresponds to the low-lying inside excitations, 
and find under which conditions inequality (\ref{STable}) is valid. 

It turns out the analog of Landau criterion for  Eq. (\ref{STable}) 
is rather simple \cite{RL}
\begin{equation}
\vert \tilde{\mu}\vert (N_{\rm o},\,v)  >  
\mu_{\rm cr} \simeq  {\rm const}\,(2\,m_{\rm x}c_{\rm s}^{2}), 
\label{cr_mu}
\end{equation}
where  $m_{\rm x} \simeq 2.7\,m_{\rm e}$,  
$c_{\rm s} \simeq  4.5 \cdot 10^5\,\,{\rm cm}/{\rm s}$.
For $z \simeq  3 \sim 10$, we estimate ${\rm const}(z)$ in Eq. (\ref{cr_mu})  
as of the order of $10^{-1}$, and we obtain the following result  
\begin{equation}
\mu_{\rm cr} \simeq 10^{-5} - 10^{-4}\,{\rm eV} 
\simeq (10^{-4} - 10^{-3})\,
E_{\rm x}. 
\end{equation}
Qualitatively, inequality (\ref{cr_mu}) is valid for the relatively high velocities 
and  numbers of the condensed particles.
In theory, one can fix the number 
$\bar{n}_{\rm o}=N^{*}_{\rm o}/N_{\rm o}$ and obtain an analog of Landau critical velocity, 
but $v_{\rm cr} < v < c_{\rm s}$ is the stability criterion  in this case.    

However, within approximation (\ref{mu}),(\ref{estimate1}) and with 
$\vert\tilde{\varepsilon}_{0}(v/c_{\rm s})\vert \simeq 10^{-1}\,E_{\rm x}$ 
\,and\,  $\bar{n}_{\rm o} \simeq 10^{1}$, 
we have  \,$\vert \mu \vert  \simeq \mu_{\rm cr}$ \,and\, 
$L_{\rm ch} \simeq 4\cdot 10^{2}\,a_{\rm x} \simeq (2 - 3) \cdot 10^{3}$\,{\AA}. 
We speculate that such a ballistic condensate can be considered as a stable 
one in the limit of $T\rightarrow 0$, see Fig. 1.  
For comparison, the condensates with 
$\vert\tilde{\varepsilon}_{0}\vert \simeq 10^{-2}\,E_{\rm x}$ and 
$\bar{n}_{\rm o} \ge 50 - 100$ 
seems to be unstable against the inside excitations. If this is the case, the continuous
generation of the low-lying inside excitation takes place,  
$$
N_{\rm o} \rightarrow N_{\rm o} - \delta N(t),
\,\,\,\,\,\partial_{t}\,\delta N_{\rm in-ex}(t) > 0, \,\,\,\,\, \
\delta N_{\rm in-ex}(t) > \sqrt{ N_{\rm o} }. 
$$ 
This process can be accompanied by the emission of outside excitations as well.  

To conclude, more accurate investigation of the stability problem is necessary. 
Indeed, the characteristic width of the  quasi-stationary solution near 
the threshold 
of stability is of  $(1 - 3) \cdot 10^{3}$\,{\AA}, whereas the typical length of the 
crystal used for experiments  is of $(2 - 4) \cdot 10^{-1}$\,cm.
We speculate that the ballistic (superfluid) propagation of 
the condensate is more than changing $x \rightarrow x-vt$
and adding $\exp(i k_{0} x)$ to the solution of
Eqs. (\ref{1Deq}) and (\ref{1Deq1}).
For example,  
the localized solitonic solution (\ref{ansatz1}),(\ref{ansatz2})    
can be used as a reasonable initial condition for modeling of how
the coherent part of exciton-phonon packet actually moves in the presence
of thermal phonons  and/or point scattering centers, etc..   

\begin{figure}[t]
\begin{center}
\leavevmode
\epsfxsize=29pc % will enlarge or reduce the postscript figures based on the xsize
\epsfysize=19pc
\epsfbox{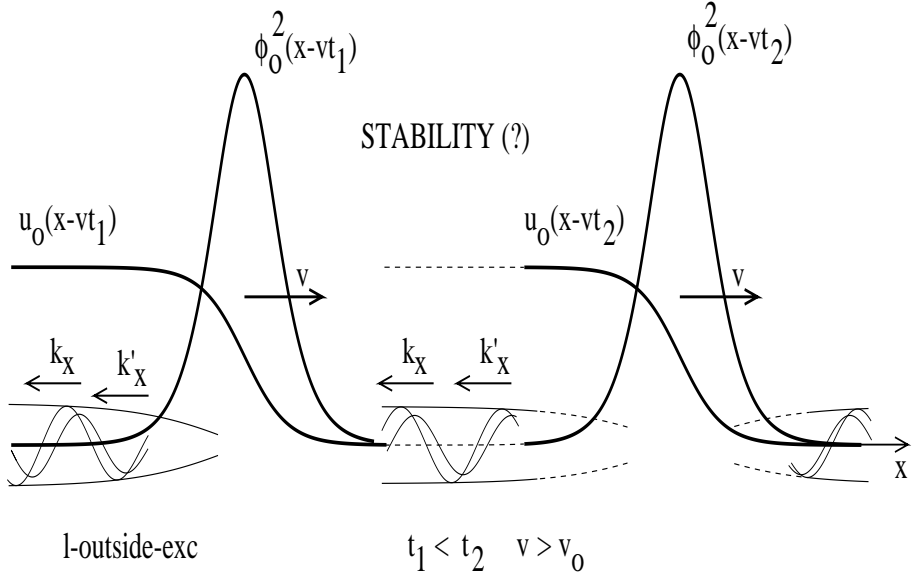} % postscript image file name
\end{center}
\caption
{The ballistic condensate,
 $\phi_{\rm o}(x-vt)\cdot u_{\rm o}(x-vt)\delta_{1j}$,
seems to be stable in relation to 
the direct emission of  
outside exciton-phonon excitations. 
(We consider the backward emission in the long-wavelength limit, 
$\vert  k_{x} \vert \simeq  z\,L_{0}^{-1}$ and $z \ll 1$.)
The outside excitations presented on this figure are  labeled by the wave vectors,   
$k_{x},\,\,k'_{x} < 0$ in 
the co-moving  frame. To a first approximation, the outside excitations 
can be described  in terms of free excitons and free (acoustic) phonons emitted from
the condensate coherently. }
\label{fig:solitons}
\end{figure}
%%%%%%%%%%%%%%%%%%%%%%%%%%%%%%%%%%%%%

\section{Discussion}

Recall that  
the self-consistent exciton-phonon condensate seems 
to be only a part of the real moving packet. 
The noncoherent part of it,  the noncondensed excitons 
$\Delta n(x,t)$ and the unidirectional phonon wind   $\Delta u(x,t)$, effects the propagation
of the condensate. 
Here we address the question on  whether  it is possible to diminish
(ideally, to turn off)
the phonon wind 
after the moving exciton-phonon condensate has been formed.
As a result, the diffusive noncondensed excitons can be delayed, 
and the coherent signal and
the noncoherent one are separated in time. 

\begin{figure}[t]
\begin{center}
\leavevmode
\epsfxsize=37pc % will enlarge or reduce the postscript figures based on the xsize
\epsfysize=22pc
\epsfbox{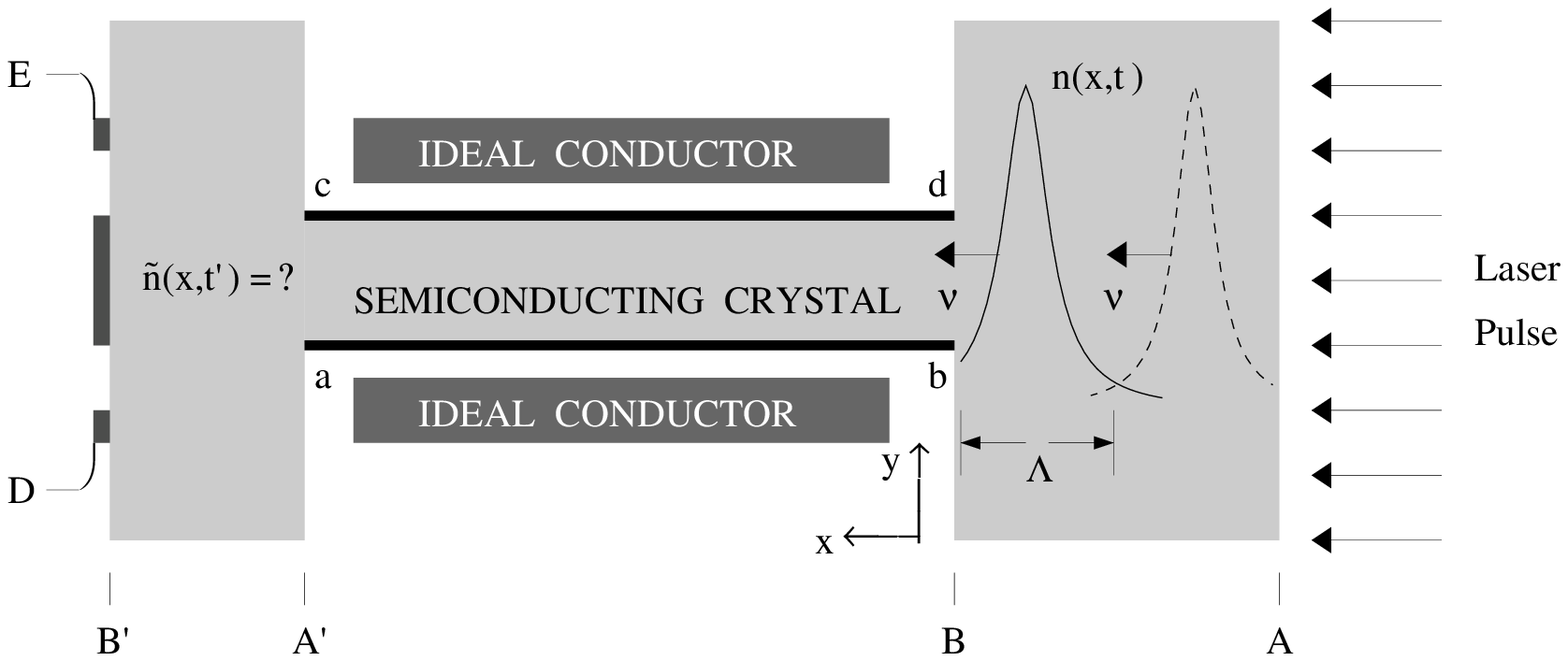} % postscript image file name 
\end{center}
\caption
{In order to separate the coherent and noncoherent components  
of the total exciton-phonon packet at $T \ll T_{\rm c}$,  it is crucial to turn off
a phonon wind. (The phonon wind, 
or a sequence of nonequilibrium acoustic phonons,  
``blows'' unidirectionally from surface A of the crystal.) 
This can happen when the packet $n(x,t)$ 
enters the channel abcd of the width $l$  
in the $y$ direction.   
The characteristic lengths of the total packet and the condensate, 
$\Lambda$ and  $L_{0}$, respectively, 
satisfy the inequality
$L_{0}< l <\Lambda$. Then, the condensate could move ballistically
through the channel with the velocity $v$, whereas the 
non-condensed excitons will diffuse (almost) freely inside the channel
because the phonon wind cannot penetrate into it. 
As a result, the (total) excitonic current 
$\tilde{n}(x,t') \approx \tilde{n}_{\rm o}(x-vt') + \Delta\tilde{n}(x,t')$ 
from the output of the channel  ($x={\rm ac}$)
can be converted into the electric current near the surface ${\rm B}'$; 
\,$\vert {\rm A}'{\rm B}' \vert  \ll \vert {\rm A}{\rm B}\vert $.
%%%%%%%%%%%%
%%% The presented apparatus is, in fact, a SEPARATOR
%%%%%%%%%%%%
}
\label{fig:separator}
\end{figure}

We consider a system consisting of a crystal (semiconductor) and two ideal
conductors (metals). 
The geometry of such a system is shown on Fig. \ref{fig:separator}. 
Let
$ l=\left|{\rm db}\right|= \left|{\rm ac}\right|$ be the distance between 
two conducting planes
surrounding part ${\rm A'B}$ of the crystal. 
Let $\lambda$ be a wavelength of
the light emitted during the exciton recombination. (Usually, $\lambda$
lies in the visible wavelength area, $\simeq 500$\,nm.)
If the following relation 
$$
l \le \lambda \ll D
$$
can be realized by  appropriate changes in the  growing 
technique, an interesting physics of penetration into a channel 
appears. 
Here  $D \simeq 1 - 2$\,mm  is the diameter of an excitation spot on 
the  surface ${\rm A}$ of the crystal, $D\simeq \sqrt{S}$.

\begin{figure}
\begin{center}
\leavevmode
\epsfxsize=37pc 
\epsfysize=22pc
\epsfbox{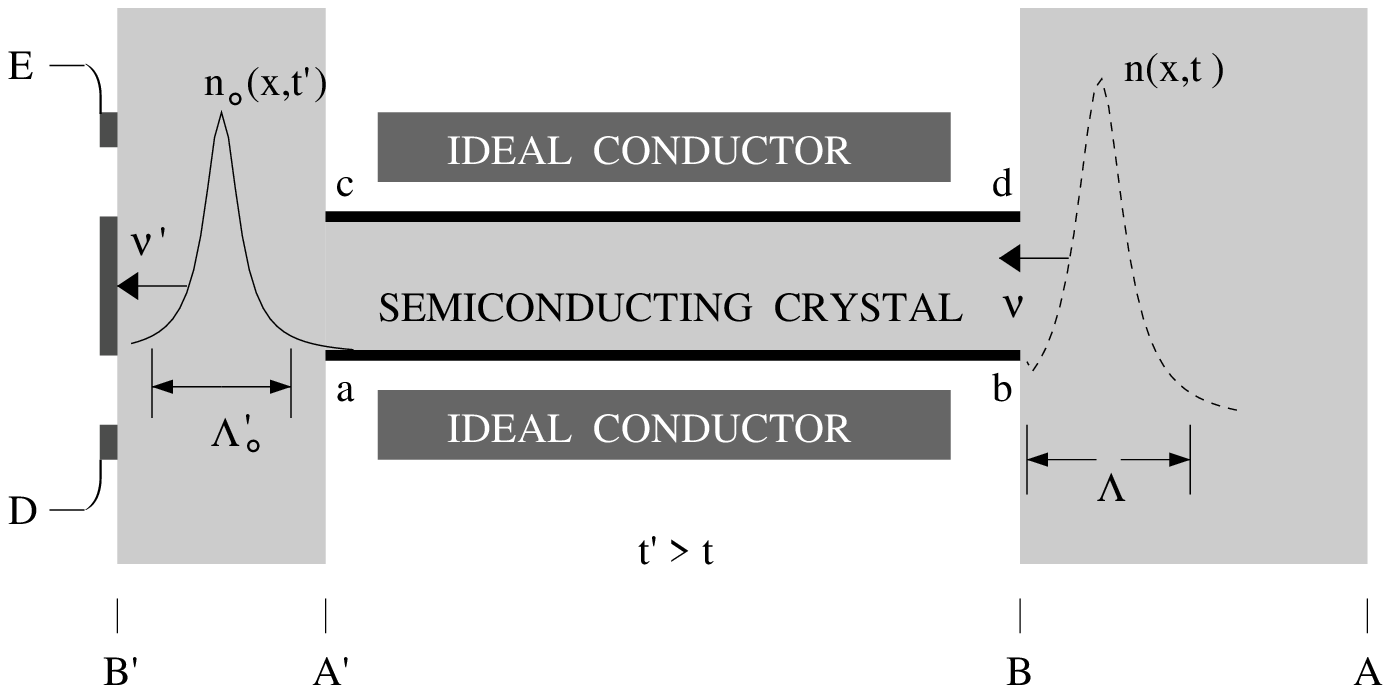} % postscript image file name 
\end{center}
\caption
{
We assume that  
the transition from the 3D crystal AB 
to the channel abcd can be done smoothly enough 
in order not to create turbulence in the condensate
penetrating the channel.
Then, one can 
expect the solitonic shape of the coherent part of 
the droplet at the end of the channel.
(The noncoherent part 
at $t=t'$ is not shown on this figure.) 
}
\label{fig:separator_at_work}
\end{figure}

Note that 
due  to boundary conditions on the conducting walls, 
it is impossible to emit
light with the wave vector less than 
$\pi/l$ and the quantum energy less than $\hbar c\, \pi/l$.
If the  last value  is 
larger than the energy of the electron-hole recombination,   the later is
suppressed in the region ${\rm abcd}$. Then the lifetime of excitons  
(both free and bound ones) can be  increased by several
orders of magnitude.

We are interested in evolution of the exciton-phonon system when the cloud 
reaches $x={\rm B}$ (see Fig. 2), 
and the excitons begin to move in channel ${\rm abcd}$. 
Roughly speaking, 
we have such a  connection problem, in which the boundary
conditions and the characteristic length scales  ($L_{y}$ in our case) can be  quite
different for the 3D crystal and  the channel, respectively. 
As boundaries $\rm ab$ and $\rm cd$ 
of the channel  are metallized, the boundary 
conditions of the displacement  
field ${\bf u}(x,y,t)\approx (u_{x},0,0)$ 
can be taken  zero along them.

We consider  transition of the phonon wind from  
part ${\rm AB}$ of the crystal to the channel   
by posing of the time-dependent boundary condition 
at the connection ${\rm B}\, \vert\,{\rm bd}$ (see Fig. 2): 
\begin{equation}
\partial_{x} u (x={\rm B},y,t) =f(y,t) \sim 
\int_{-\infty}^{\infty} \frac{ {\rm d}\omega}{2\pi} \,
{\rm e}^{-i\omega t}\, f(\omega),  
\label{force}
\end{equation}
where $f(\omega)$ is  localized  inside $\vert  \omega \vert < \omega^{*}$.
For the displacement field  inside the channel,  
we can write the following representation,  
\begin{equation}
u(x,y,t)=\sum_{n = {\rm odd} }^\infty u_n(x,t)\,\cos\frac{\pi n y}{l}, \,\,\,\,\,\vert y \vert <l/2, \,\,\,\,\,
u_n(x,t) =  \int_{-\infty}^{\infty} \frac{ {\rm d}\omega }{2\pi}\, \tilde{A}_{n}(\omega )\,{\rm e}^{ik(\omega)x-i\omega t}.
\label{field1}
\end{equation}
Obviously,  all the  modes in Eq. (\ref{field1}) 
with frequencies less than $\omega_0 \simeq c_{\rm s}\pi /l$ 
are dumped inside the  channel.
Let the  following inequality be  valid
$$
\omega^{*} \simeq \frac{2\pi}{\tau }\simeq \frac{2\pi c_{\rm s} }{\Lambda } < (\ll)\, \omega_0 \simeq 
\frac{\pi c_{\rm s}}{l}.
$$
Here, $\tau$  and $\Lambda$ can be estimated from the duration and  the characteristic width 
of the (total) exciton-phonon packet driven  by the phonon wind, 
$\tau \simeq (1\sim 5)\cdot 10^{-7}$\,s and  
$\Lambda \simeq 10^{-2}$\,cm.
Then, only the ``tails'' of $f(\omega)$ are actually transfered into
the amplitudes $\tilde{A}_{n}(\omega )$ of the moving packet $u_{n}(x,t)$. 
Hence, 
the phonon drag force ($\propto \partial_{x} u(x,y,t)$) can be  essentially suppressed
inside the channel, 
and  the non-condensed excitons can diffuse almost freely there.

We speculate that the exciton-phonon condensate being formed in AB part of the 3D crystal
can, first, penetrate into and, second,  
pass through the channel (see Fig. 3). 
This problem seems to be not a quasi-stationary one. 
Indeed, within approximation $\Psi_{0}=\psi_{\rm o}(x,y,t)$ \,and\, 
${\bf u}_{0} = (u_{{\rm o}\,x}(x,y,t),\,0,\,0)$, we have the following system  to describe 
the dynamics at $x>{\rm B}$: 
%% $$
%%  i\hbar\partial_{t}{\psi}_{\rm o}(x,y,t)= 
%% $$
\begin{equation}
i\hbar\partial_{t}{\psi}_{\rm o}(x,y,t)= 
\Bigr(- {{\hbar^{2}}\over{2m_{\rm x}}}\Delta_{x,y} + \nu_{0}
\vert{\psi_{\rm o}}\vert^{2}(x,y,t)+ 
\nu_{1}\vert{\psi}_{\rm o}\vert^{4}(x,y,t)\,\Bigl)  
{\psi}_{\rm o}(x,y,t)\,+ 
\label{eq11}
\end{equation}
$$
+\,\sigma_{0}\,\partial_{x}{ u}_{{\rm o}\,x}(x,y,t)\,
{\psi}_{\rm o}(x,y,t),
$$
%%%%%%%%%%%%%%%%
%
\vspace*{0.1cm}
%
%%%%%%%%%%%%%%%%%
%%$$
%% \bigl(\partial_{t}^{2}- c_{\rm s}^{2}\Delta_{x,y} \bigr){u}_{{\rm o}\,x}(x,y,t)
%% -c_{\rm s}^{2}  \sum_{j=x,y}\!
%% 2 \kappa_{3}\,\partial_{j}^{2} { u}_{{\rm o}\,x}\,\partial_{j}{ u}_{{\rm o}\,x}(x,y,t)
%% \,=
%%% $$
\begin{equation} 
\bigl(\partial_{t}^{2}- c_{\rm s}^{2}\Delta_{x,y} \bigr){u}_{{\rm o}\,x}(x,y,t)
-c_{\rm s}^{2}  \sum_{j=x,y}\!
2 \kappa_{3}\,\partial_{j}^{2} { u}_{{\rm o}\,x}\,\partial_{j}{ u}_{{\rm o}\,x}(x,y,t)
=
\rho^{-1}\sigma_{0}\,\partial_{x}\bigl(\vert\psi_{\rm o}\vert^{2}(x,y,t)\bigr),
\label{eq22}
\end{equation}
%%%%%%%%%%
%
\vspace*{0.1 cm}
%
%%%%%%%%%%%%%%%
\begin{equation}
{\psi}_{\rm
o}(x,-l/2,t)={\psi}_{\rm o}(x,l/2,t)=0,\,\,\,\,\,{u}_{{\rm
o}\,x}(x,-l/2,t)={u}_{{\rm o}\,x}(x,l/2,t)=0.
\label{eq22B}
\end{equation}
Boundary conditions (\ref{eq22B}) model the metallized surfaces of the channel. 

Here, we  model a kind  of  smooth penetration, for example,  
no vortexes are generated 
near the 3D-channel intersection. 
Recall that the  width of the condensate  in our quasi-1D model
is  $L_{0} \simeq (10^{2} - 10^{3})\,a_{\rm x} < \Lambda$,
and \,$\tau_{0} \simeq  L_{0}/c_{s} \simeq 10^{-10} -  10^{-9}$\,s\,$< \tau$.
As \,$L_{0} < l < \Lambda $  and  \,$\omega^{*} <  \omega_{0} < \tau_{0}^{-1}$,
we assume that the localized nonlinear condensate can  penetrate
into the beginning of the channel without the loss of its coherence. 
Then, instead of solving Eq. (\ref{eq11})-(\ref{eq22B}) with  ``pumping'',    
(i.e., time dependent) boundary conditions at $x={\rm B}$,
one can transform the quasi-stationary solution  
(\ref{ansatz1}) and (\ref{ansatz2})
into  the  initial condition inside the channel and choose $t=0$.
For example, one can use the following trial functions:  
$$
\psi_{\rm o}(x,y,0)\cdot u_{\rm o}(x,y,0)\,\delta_{1j}\,=
$$
%%%%%%%%%%%%%%%%%
\begin{equation}
=\, \exp(im_{\rm x}vx)\,
\Phi_{\rm o}\cosh^{-1}\bigl( L_{0}^{-1}x\,\bigr )\,\bar{\phi}(y) \cdot
\Bigl( Q_{\rm o}\,-\,Q_{\rm o}\tanh \bigl(L_{0}^{-1}x\bigr)\, \Bigr )\,\bar{Q}(y), 
\label{movingcond2}
\end{equation}
where $\bar{\phi}(y)$ and $\bar{Q}(y)$ 
have to satisfy the boundary conditions (\ref{eq22B})
and be self-consistent as well.
We assume that near the boundaries 
they have the same coherence length $L_{0\,y}$. It can be estimated as  \cite{Fetter}
\begin{equation}
L_{0\,y}^{-1} \simeq \sqrt{\nu_{0}\,\vert \phi_{\rm o}(x)\vert^{2}\,m_{\rm x}  }/\hbar
\,\rightarrow\, 
\sqrt{(2m_{\rm x}  /\hbar^{2})\,\nu_{0}\,\Phi_{\rm o}^{2}/2 }\,\cosh^{-1}\bigl( L_{0}^{-1}x\,\bigr ).
\label{healingL}
\end{equation}
Near the soliton maximum ($x=0$),  we obtain   $L_{0\,y} \simeq L_{0}$
(see Eq. (\ref{estimate1}) and 
$\nu_{0} \simeq  \vert \tilde{\nu}_{0} \vert $ is valid in the stability area).

We estimate that as many as 
$\simeq  N_{\rm o}\,(l/D) $ \,of\, the $N_{\rm o}$ original 
Bose condensed excitons   
can penetrate from  the crystal into  the channel  
if only $l/2 > L_{0\,y}(x)$ \,for\,  $\vert  x \vert < (2\sim 4) L_{0}$.  
Then, the BEC conditions can be saved  for the droplet inside the channel
at  $T\ne 0$,  
and the exciton-phonon condensate can move through the
channel {\it ballistically} (Fig. 3). 
However, the question of stability 
of such a motion remains to be open \cite{Rou}.
Experimentally, 
the task is to register the signal from the output of 
the channel, $x={\rm ac}$. 
For example, a small 3D part could  be attached to the  channel in order to 
convert the excitonic mass current 
into a measurable electric signal, see Figs. 2 and 3.

\section{Conclusion}

In conclusion, we note that  
the phonons play a crucial role 
in almost all the current models aimed to explain or predict  
coherent behavior of excitons in semiconductors, see, f.ex., 
\cite{Lozovik},\cite{AIm},\cite{Ivanov}.
Although our model of the exciton-phonon condensate 
fails to predict the width of the exciton-phonon packet correctly,  
the theory yields a qualitative description of the experiments and
a reasonable value for the critical velocity. 
However,  one has to take into account the thermal excitons 
(e.g., the weak tail that is always observed behind the soliton \cite{Andre}) 
and the thermal phonons of the crystal 
to make the model {\it with condensate} more realistic.

\end{document}